%% file: main.tex
\documentclass{easychair}

\usepackage{array}
\usepackage{enumerate}
\usepackage{tikz}
\usepackage{graphicx}

\usepackage{colortbl}
\usepackage{xcolor}

\usepackage{xspace}

\usepackage{gensymb}

\usepackage{listings}

\usepackage{comment}
\usepackage{multirow}
\usepackage{amssymb,amsmath}
\usepackage{booktabs}

\title{Formal Verification of a Programmable Hypersurface
	\thanks{This work was partially funded by the European
	Union via the Horizon 2020: Future Emerging Topics
	call (FETOPEN), grant EU736876, project VISORSURF
	(http://www.visorsurf.eu).}
}
\author{Panagiotis Kouvaros\inst{1}
	\and  Dimitrios Kouzapas\inst{1}
	\and Anna Philippou\inst{1}
	\and Julius Georgiou\inst{2} 
	\and Loukas Petrou\inst{2}
	\and Andreas Pitsillides\inst{1}
}
\institute{
	Department of Computer Science, University of Cyprus\\
	\email{\{pkouva01,dkouza01,annap,cspitsil\}@cs.ucy.ac.cy}
	\and
	Department of Electrical and Computer Engineering, University of Cyprus\\ 
	\email{\{julio,lpetro02\}@ucy.ac.cy}
}

\newcounter{noteCounter}

\begin{document}
	\maketitle

	\input{abstract}

	\pagestyle{plain}
	\pagenumbering{arabic}
	
	\input{intro}

	\input{hypersurface}

	\input{evaluation}

	\input{conclusions}

	\bibliographystyle{abbrv}
	\bibliography{references}
\end{document}

%% file: abstract.tex
\begin{abstract}
	A metasurface is a surface that consists of artificial
	material, called metamaterial, with configurable
	electromagnetic properties.
	This paper presents work in progress on the design and
	formal verification of a programmable metasurface, the Hypersurface,
	as part of the requirements of the VISORSURF research program
	(HORIZON 2020 FET-OPEN).
	The Hypersurface design is concerned with the development
	of a network of switch controllers that are responsible
	for configuring the metamaterial.
	The design of the Hypersurface, however, has demanding
	requirements that need to be delivered within a context
	of limited resources.
	This paper shares the experience of a rigorous design
	procedure for the Hypersurface network,
	that involves iterations between designing
	a network and its protocols and the formal evaluation
	of each design.
	Formal evaluation has provided results that,
	so far, drive the development team in 
	a more robust design and overall aid in reducing the
	cost of the Hypersurface manufacturing.
\end{abstract}

%% file: intro.tex
\section{Introduction}

This paper reports on work-in-progress
carried out  in the context of the research programme
``VISORSURF: A Hardware Platform for Software-driven Functional
Metasurfaces''~\cite{visorsuf_webpage}, funded by Horizon 2020
FET-OPEN. 
VISORSURF is an interdisciplinary programme between  computer
science (networks/nano-networks and formal methods), computer
engineering (circuit design and implementation), and physics
(meta-materials). Its main objective is to develop a hardware
platform, the {\em HyperSurface} (HSF), whose electromagnetic
behavior can be defined programmatically.
The HSF's enabling technology are  {\em metasurfaces}, artificial
materials whose electromagnetic properties depend on their internal
structure. Controlling the HSF is a network of controller switches
which receives external software commands and alters the metasurface
structure yielding a desired electromagnetic behavior, thus allowing
a number of high-impact applications. These include electromagnetic invisibility 
of objects, filtering and steering of light and sound, as well as
ultra-efficient antennas for sensors and communication devices. 

This paper is concerned with the requirement of the programme for the
rigorous design and formal evaluation of the controller-switches network
and its protocols. This requirement stems from the project's 
challenge to provide cutting-edge technology with limitations in both
time and cost.
Indeed, it is of paramount importance for the produced hardware to adhere
to its specification from the very first version of the product, given the high
cost of producing the components and the fact that
the project's budget is fixed.
The specification includes qualitative properties, e.g., the controller
network should route all messages correctly to all network nodes, as well
as quantitative properties, since  nodes need to be reached within specified
time bounds in a fault-tolerant manner while preserving power. 

Typically, in the
networks literature, evaluation of network topologies and protocols is carried out
via extensive simulation using discrete-event simulators such as NS-2 or
OPNET, or via testbed experiments. While these are important 
evaluation methods, the results obtained are highly dependent on the 
physical-layer models supported by the simulators and, in the case of
experiments, they  are not suitable during the design phase
of a protocol. At
the same time, as is well known,
the simulative approach may discover flaws in a system but it cannot prove their absence. 
On the other hand, formal analysis techniques allow to formally verify
that a system complies to its specifications and check for the absence of flaws. 
Model checking, in particular, allows to investigate the behavior of a
model via an exhaustive search of its state space. Properties of interest
may be enunciated in temporal logic and subsequently checked for 
satisfaction on all possible executions of the system. In case of property
violation, counter-examples can be provided to support the designer to 
diagnose the error. Model checking has been applied for the analysis and design
of network protocols in a number of works including~\cite{BhargavanOG02,FehnkerGHMPT12,DombrowskiJKG16}.

Unfortunately, a main drawback associated with model checking is the state-space explosion problem
and on many occasions analysis cannot be
applied on systems of a realistic size. To this effect, the use
of statistical model checking (SMC) has been advocated.
Statistical model checking~\cite{SenVA05,PhDYounes} is a formal-analysis approach that combines
ideas of model checking and simulation with the aim of supporting quantitative analysis
as well as addressing the state-space explosion problem. It uses Monte Carlo style sampling 
and hypothesis testing to provide evidence that a system satisfies a given property with
high probability. The main idea is to simulate the system for finitely many
runs and use hypothesis testing to infer whether the samples provide a statistical evidence for the satisfaction
or violation of the specification. Naturally, the greater the number of simulations, the higher the precision
achieved. The benefits of employing statistical model checking towards the analysis of network algorithms
have been illustrated in various works including~\cite{HofnerK13,HofnerM13,CorsoMM15}.

As required by the project's objectives, our goal has been to develop a set of 
network protocols (network initialisation, routing and reporting) on 
a grid network (a Manhattan style topology~\cite{MSN}
imposed by the hardware requirements of the project). In this paper
we focus on the design of the routing algorithm, which proved to be the
main challenge of the work. 
Routing within grid networks has been a topic of thorough investigation within the
network community and it has been of great interest in domains such as networks on
chip~\cite{DallyTowles,NOCs}. Various algorithms have been proposed in the 
literature for 
mesh topologies where the main challenges posed were towards providing efficiency and
tolerance to faults~\cite{XY-on-NOCs,DyXY,XYX,Odd-even}.
 While these works influenced the development of our
routing protocol, the various restrictions imposed by the specific application, such as
the limited connectivity as well as the limited resources available
to each network node (e.g. limited memory/buffering space, limited computational capabilities)
rendered the design of the routing algorithm quite challenging.  Indeed, 
it turns out that apparently innocent  characteristics of
our model (e.g. the lack of line/column wrap-arounds) create the risk of deadlocks,
even in the absence of faults in the network. To address this problem
(discovered via model checking), it was necessary to explore
options such as introducing buffers in the nodes or adopting different routing sequences so as to
handle the congestion of parts of the network, and to seek methods for assessing these options and
provide guarantees that they satisfy the set requirements.

Taking the requirements of VISORSURF into account has led us to employ formal methods
from the initial stages of the iterative design of the network protocols via a
continuous assessment of design proposals against requirements using
model checking.
Early on, our  experimentation confirmed
that the state-space explosion problem
is a severe limitation when attempting to analyse a network of a reasonable size.
Thus, we turned towards Statistical Model Checking (SMC) and we employed
the UPPAAL tool and, more specifically, its SMC extension~\cite{UPPAAL}.
In this paper we report on our experience of applying formal methods in the design
phase of a routing algorithm on a grid network as
imposed by the hardware requirements of the project, and how this led to important design
decisions, thus significantly facilitating us towards our goal. 
Furthermore, we discuss the main challenges we faced in obtaining
desired results which point out directions for further research.
We believe that our conclusions provide evidence on the impact
formal methods may have in the design and implementation of
technological applications in the context of small and medium-scale projects.

%% file: hypersurface.tex
\section{The Hypersurface: Requirements and Design Parameters}
\label{sec:hypersurface}

In this section we present the main requirements and design parameters
of the Hypersurface, as imposed in the context of the VISORSURF
programme and as needed in the present discussion. We identify three levels of requirements:
i) architectural/physical constraints as imposed by the physical level of
the HSF;
ii) VISORSURF programme requirements as approved by the funding authority;
and 
iii) resource/manufacturability limitations, in both time and money
 that make the design phase a non-trivial task.

\paragraph{Architectural/Physical Constraints and Terminology.}
The {\em metasurface tile} is a surface consisting of
configurable {meta-material} strips arranged as a grid. A set of four meta-material
strips is configured via a {controller switch}, also called the
{\em intra-tile controller}.
All intra-tile controllers of the HSF are interconnected to
constitute the {\em intra-tile network}.
Intra-tile controllers will be designed and
implemented as a single hardware element 
and their purpose is to implement basic functionalities, most importantly, support the rudimentary
routing of {\em configuration packets} for configuring the
metamaterial.

The intra-tile network  receives configuration data from one or more
{\em gateway controllers}. A gateway controller resides on
the periphery of the metasurface and it sends configuration
packets to  controllers throughout the network that, in turn, are programmed by the user.
A gateway controller has full computing power. 
It is envisaged that  tiles will be interconnected
at the gateway controller level to form larger metasurfaces.

\paragraph{VISORSURF Requirements.}
As already explained, an intra-tile controller's main task is to set the EM properties of the
meta-material strips as directed via configuration packets from the gateway.
Note that these packets are directives for appropriately implementing a desired
functionality (e.g., to absorb or steer impinging waves) and, for any given function,
they consist of one message per network controller. Such a set of configuration
packets can be delivered in any order, thus allowing the flexibility
to the gateway to decide on the sequence in which the packets will be delivered to
the controller nodes. We refer to such sequences as  {\em configuration sequences}.

In addition,
intra-tile controllers are expected
to report acknowledgements and status to the gateway,
thus enabling the monitoring of the state of the controller network in real time and
hence ``debug'' the HyperSurface program.
As such, the intra-tile controller network needs to implement  routing for both data
and acknowledgement packets.
The routing
should be flexible, scalable, and robust. Furthermore,  packets should be delivered 
in a timely manner (where the timing constraints will be determined in the course
of the project). 
Finally, the intra-tile network needs to provide mechanisms that support a high
degree of fault tolerance, where data packets will continue to be delivered
to the recipient controllers despite hardware faults.

\paragraph{Resource/Manufacturability Limitations.}
The programme is required to deliver a functioning HSF prototype
within a specific amount of time, money, human, and expertise resources.

The main hardware element to be manufactured is the
intra-tile controller. To limit the overall cost, a single uniform type
of controller will be designed and manufactured. 
The selected chip technology for the controller manufacturing allows for
a maximum number of 25 pins per intra-tile controller chip. The restriction
limits the interconnection capabilities of an intra-tile
controller with other components of the metasurface such as
its connectivity with its neighbouring controllers as well as with the gateway.
A consequence of this restriction is that intra-tile controllers will transmit data in 
a single bit-by-bit scheme.
Moreover, this communication will be implemented asynchronously via an appropriate 
four-way asynchronous communication hardware protocol.
Asynchronous communication uses no clock for synchronisation. Instead, the sender
relies on the acknowledgement signal of the receiver to start and end a transmission.
The restriction of asynchronous communication was imposed since 
adding a clock to the chip of the controller would have the following undesirable implications:
i) require more components, such as a crystal that will increase the chip size,
and a phase-lock loop responsible for inter-controller synchronisation;
ii) increase power consumption; and
iii) make a total metasurface absorber impossible because of the clock's
electromagnetic emissions.
Finally, we mention that intra-tile chips will only possess volatile memory
since non-volatile memory is expensive and error-prone.

\subsection{Hypersurface Manufacturing: Iteration-$0$}

In order to mitigate 
the implementation risk, manufacturing of the intra-tile chip will take place in
iterations. The first manufacturing iteration is expected to
implement a basic but working prototype, and the entire design process will
be completed for the final
deliverable.

The experience presented in this paper will be implemented in the
first manufacturing iteration: iteration-$0$. Despite its basic
functionality, iteration-$0$ identifies the elements
that are going to be used by all future iterations:
controller hardware and communication protocols, controller
pin allocation, network topology, packet format, basic extendable
routing protocol, and basic functionalities.

\input{fig-pin_allocation}
The initial design for iteration-$0$ can be found in
Fig.~\ref{fig:pin_allocation} and Fig.~\ref{fig:network}.
The three diagrams in Fig.~\ref{fig:pin_allocation} demonstrate
the allocation of the pins and the communication channel endpoints on
the intra-tile controller chip. Each channel endpoint
requires three pins to implement bit-by-bit asynchronous
communication. The limited number of pins (25) limits
to a design where only four unidirectional channel
endpoints can be allocated (a total of 12 pins) per
controller. The physical distribution of the pins is as in
diagram (b).

\input{fig-network}

Following the design of the intra-tile controller,
the suggestion for a grid topology is a variation of the Manhattan 
network topology~\cite{MSN} as presented in 
Fig.~\ref{fig:network}. Its main characteristic is that the routing direction alternates at each consecutive row and column. 
The topology is achieved by rotating the single design 
intra-tile controller by 90\degree~each time to get the
four different orientations (a-d) that are shown in Fig.~\ref{fig:orientation}.
The interconnection of the four orientations is used
to achieve the Manhattan topology; depending on the
physical orientation of each intra-tile controller
an output endpoint is connected to the corresponding
input endpoint of a neighboring intra-tile controller.
Each intra-tile controller has knowledge about its
type based on its address.

\input{fig-orientation}

The proposed topology offers a flexible and robust network, which
respects the design constraints:
it provides connectivity between the network nodes 
using only two input and two output edges per node.
Unlike the Manhattan networks considered in the literature,
the proposed topology provides connections (and consequently
bidirectional communication) between neighbouring
periphery nodes, which we refer to as {\em wrap-arounds},
thus employing all communication channels of the nodes and
providing connectivity between all nodes. 
Our design choice of connecting neighboring periphery nodes
and not the ends of each row and each column is due to
the hardware implementation: crossing the 
interconnection wires would require to add extra layers on the PCB board
that embeds the meta-surface. Furthermore, the edge controllers  
would require components, e.g. transistors, with more signal drive to send 
signals over longer wires.

Moving now to the programming of the chip, we point out that
there are two modes of operation: the initialisation mode
and the normal operation mode. This paper is concerned
with evaluating the normal operation mode.
The initialisation mode is used to initialise
each intra-tile controller with a unique address and
with additional initialisation data. This is necessary since,
as already discussed, only a single type of controller will
be produced and will not possess any non-volatile memory.
This has led
to the design of a simple initialisation protocol that will assign
an address to each controller (its X-Y coordinates), which will be stored at its volatile
memory, and, based on which each controller will determine
its ``type'' based on its coordinates.

In the normal operation mode, due to the
limited computing power of the intra-tile controller, we are
experimenting with variants of the simple XY routing
protocol~\cite{XY-on-NOCs}, adopted for the Manhattan topology.
Below there is the 
simple XY protocol variant adopted for the iteration-$0$ design.
\begin{table}
\begin{lstlisting}
XY routing algorithm(packet)
	x, y: address a, b: target address

(a, b) = packet
  if(x == a) {
  	if(y == b)
   		send ack;
	else if (y < b)
		send packet up
	else if (y > b)
		send packet down
  }
else if(x < a - 1)
	send packet right
else if (x == a - 1) {
	if(x mod 2 = 0 and y < b)
		send packet up
	else
		send packet right
}
else if(x > a) {
	if(y == b)
		send packet left
	else if(y < b)
		send packet up
	else if(y > b)
		send packet left
}
\end{lstlisting}
\caption{Pseudocode for the XY routing protocol variant}
\end{table}
The XY
routing protocol assumes a Cartesian coordination system at
the intra-tile controllers grid. 
The implementation assumes a gateway controller connected at the south west
corner of the network grid and sending routing packets to intra-tile
controller $(0, 0)$.
The protocol first routes a packet on the $x$-axis until it reaches the
target $x$-coordinate and then similarly on the $y$-axis until
it reaches the target. In a Manhattan topology we assume a standard
mapping of the four directions ``up'', ``down'', ``left'', ``right''
on each intra-tile controller depending on its orientation.
%
%
Upon receiving a configuration packet, an intra-tile controller
creates an acknowledgement packet to be routed to a gateway controller.

The development of the iteration-$0$ design has undergone several cycles
between design and analysis. The parameters considered at each
iteration include the number and position of the gateway controllers,
the presence of buffer space to store received packets at each intra-tile controller
as well as the capability of the controllers for parallel processing/routing of packets.
The next section describe the model and the evaluation of each
design following the design parameters of the topology.

%% file: fig-pin_allocation.tex
\begin{figure}
\vspace{-5mm}
	\begin{tabular}{l}
		\begin{tabular}{ccccc}
			\begin{tikzpicture}
				\draw		(0, 0) -- (0, 0);
				\draw[rounded corners]		(0.5, 0.5) rectangle (1.5, 1.5);
				\draw[fill]	(0.6, 0.6) circle (0.02);
				\draw[fill]	(0.8, 0.6) circle (0.02);
				\draw[fill]	(1, 0.6) circle (0.02);
				\draw[fill]	(1.2, 0.6) circle (0.02);
				\draw[fill]	(1.4, 0.6) circle (0.02);

				\draw[fill]	(0.6, 0.8) circle (0.02);
				\draw[fill]	(0.8, 0.8) circle (0.02);
				\draw[fill]	(1, 0.8) circle (0.02);
				\draw[fill]	(1.2, 0.8) circle (0.02);
				\draw[fill]	(1.4, 0.8) circle (0.02);

				\draw[fill]	(0.6, 1) circle (0.02);
				\draw[fill]	(0.8, 1) circle (0.02);
				\draw[fill]	(1, 1) circle (0.02);
				\draw[fill]	(1.2, 1) circle (0.02);
				\draw[fill]	(1.4, 1) circle (0.02);

				\draw[fill]	(0.6, 1.2) circle (0.02);
				\draw[fill]	(0.8, 1.2) circle (0.02);
				\draw[fill]	(1, 1.2) circle (0.02);
				\draw[fill]	(1.2, 1.2) circle (0.02);
				\draw[fill]	(1.4, 1.2) circle (0.02);

				\draw[fill]	(0.6, 1.4) circle (0.02);
				\draw[fill]	(0.8, 1.4) circle (0.02);
				\draw[fill]	(1, 1.4) circle (0.02);
				\draw[fill]	(1.2, 1.4) circle (0.02);
				\draw[fill]	(1.4, 1.4) circle (0.02);
			\end{tikzpicture}
			&\qquad \qquad&
			\begin{tikzpicture}
				\draw[rounded corners]		(0.5, 0.5) rectangle (1.5, 1.5);
				\draw[->]	(0, 1) node[above] {\tiny input2} -- (0.5, 1);
				\draw[->]	(1.5, 1) -- (2, 1)  node[below] {\tiny output2} ;
				\draw[->]	(1, 0)  -- node[left] {\tiny input1} (1, 0.5);
				\draw[->]	(1, 1.5) -- node[right] {\tiny output1} (1, 2);
			\end{tikzpicture}
			&\qquad \qquad&
			\begin{tikzpicture}
				\draw		(0, 0) -- (0, 0);
				\draw[rounded corners]		(0.5, 0.5) rectangle (1.5, 1.5);
				\draw[fill]	(0.6, 0.6) circle (0.02);
				\draw[fill]	(0.8, 0.6) circle (0.02);
				\draw[fill]	(1, 0.6) circle (0.02);
				\draw[fill]	(1.2, 0.6) circle (0.02);
				\draw[fill]	(1.4, 0.6) circle (0.02);

				\draw[fill]	(0.6, 0.8) circle (0.02);
				\draw[fill]	(0.8, 0.8) circle (0.02);
				\draw[fill]	(1, 0.8) circle (0.02);
				\draw[fill]	(1.2, 0.8) circle (0.02);
				\draw[fill]	(1.4, 0.8) circle (0.02);

				\draw[fill]	(0.6, 1) circle (0.02);
				\draw[fill]	(0.8, 1) circle (0.02);
				\draw[fill]	(1, 1) circle (0.02);
				\draw[fill]	(1.2, 1) circle (0.02);
				\draw[fill]	(1.4, 1) circle (0.02);

				\draw[fill]	(0.6, 1.2) circle (0.02);
				\draw[fill]	(0.8, 1.2) circle (0.02);
				\draw[fill]	(1, 1.2) circle (0.02);
				\draw[fill]	(1.2, 1.2) circle (0.02);
				\draw[fill]	(1.4, 1.2) circle (0.02);

				\draw[fill]	(0.6, 1.4) circle (0.02);
				\draw[fill]	(0.8, 1.4) circle (0.02);
				\draw[fill]	(1, 1.4) circle (0.02);
				\draw[fill]	(1.2, 1.4) circle (0.02);
				\draw[fill]	(1.4, 1.4) circle (0.02);

				\draw[->]	(0, 1.2) node[left] {\tiny data bit 1} -- (0.6, 1.2);
				\draw[->]	(0, 1) node[left] {\tiny data bit 0} -- (0.6, 1);
				\draw[<-]	(0, 0.8) node[left] {\tiny bit ack} -- (0.6, 0.8);

				\draw[->]	(1.4, 1.2) -- (2, 1.2) node[right] {\tiny data bit 1} ;
				\draw[->]	(1.4, 1) -- (2, 1) node[right] {\tiny data bit 0} ;
				\draw[<-]	(1.4, 0.8) -- (2, 0.8) node[right] {\tiny bit ack} ;

				\draw[->]	(0.8, 0) -- (0.8, 0.6);
				\draw[->]	(1, 0) -- (1, 0.6);
				\draw[<-]	(1.2, 0) -- (1.2, 0.6);

				\draw[->]	(0.8, 1.4) -- (0.8, 2);
				\draw[->]	(1, 1.4) -- (1, 2);
				\draw[<-]	(1.2, 1.4) -- (1.2, 2);

			\end{tikzpicture}

			\\
			{\scriptsize (a) Chip Pins}
			&&
			{\scriptsize (b) Two inputs/outputs}
			&&
			{\scriptsize (c) Communication: Pin allocation}
		\end{tabular}
	\end{tabular}
	\caption{Pin Allocation. \label{fig:pin_allocation}}
\end{figure}
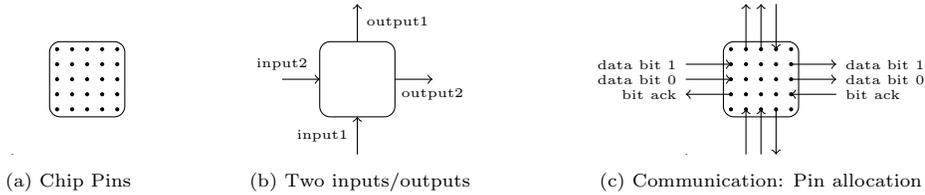

%% file: fig-network.tex
\begin{figure}
	\begin{center}
	\begin{tabular}{l}
		\begin{tabular}{c}
		\begin{tikzpicture}[scale=1]
			\foreach	\x	in	{0,...,3}
				\foreach	\y	in	{0,...,3} {
					\draw[rounded corners]	(\x, \y) rectangle (\x + 0.5, \y + 0.5);
					\node	at		(\x + 0.25, \y + 0.25) {$\scriptscriptstyle (\x, \y)$};
				}

			\foreach	\x	in	{0,...,0}
				\foreach	\y	in	{0,...,1} {
					\draw[->]	(\x-0.5, 2*\y+0.25) -- (\x, 2*\y+0.25);
					\draw[<-]	(\x-0.5, 2*\y+1.25) -- (\x, 2*\y+1.25);
					\draw[->]	(2*\y+0.25, \x-0.5) -- (2*\y+0.25, \x);
					\draw[<-]	(2*\y+1.25, \x-0.5) -- (2*\y+1.25, \x);
			}

			\foreach	\x	in	{1,...,4}
				\foreach	\y	in	{0,...,1} {
					\draw[->]	(\x-0.5, 2*\y+0.25) -- (\x, 2*\y+0.25);
					\draw[<-]	(\x-0.5, 2*\y+1.25) -- (\x, 2*\y+1.25);
					\draw[->]	(2*\y+0.25, \x-0.5) -- (2*\y+0.25, \x);
					\draw[<-]	(2*\y+1.25, \x-0.5) -- (2*\y+1.25, \x);
				}

			\draw		(0.25, -0.5) -- (1.25, -0.5);
			\draw		(2.25, -0.5) -- (3.25, -0.5);

			\draw		(0.25, 4) -- (1.25, 4);
			\draw		(2.25, 4) -- (3.25, 4);

			\draw		(-0.5, 2.25) -- (-0.5, 3.25);

			\draw		(4, 0.25) -- (4, 1.25);

			\draw[rounded corners]		(-1.1, 0) rectangle (-0.5, 1.5);
			\node[below]	at		(-0.8, 0)	{\tiny gateway};

			\draw[rounded corners]		(4, 2) rectangle (4.6, 3.5);
			\node[below]	at		(4.3, 2)	{\tiny gateway};

		\end{tikzpicture}
		\end{tabular}
	\end{tabular}
	\end{center}
	\caption{Manhattan Topology with edge wraparound \label{fig:network}}
\end{figure}
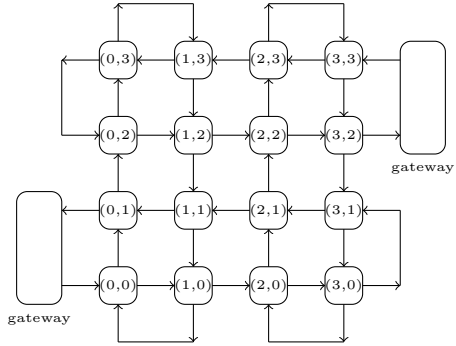

%% file: fig-orientation.tex
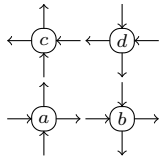
\begin{figure}
	\begin{center}
	\begin{tikzpicture}[scale=0.65]

		\draw[rounded corners]	(0, 0) rectangle (0.5, 0.5);
		\node		at	(0.25, 0.25) {\scriptsize $a$};
		\draw[->]	(-0.5, 0.25) -- (0, 0.25);
		\draw[->]	(0.5, 0.25) -- (1, 0.25);
		\draw[->]	(0.25, -0.5) -- (0.25, 0);
		\draw[->]	(0.25, 0.5) -- (0.25, 1);

		\draw[rounded corners]	(1.6, 0) rectangle (2.1, 0.5);
		\node		at	(1.85, 0.25) {\scriptsize $b$};
		\draw[->]	(1.1, 0.25) -- (1.6, 0.25);
		\draw[->]	(2.1, 0.25) -- (2.6, 0.25);

		\draw[<-]	(1.85, -0.5) -- (1.85, 0);
		\draw[<-]	(1.85, 0.5) -- (1.85, 1);

		\draw[rounded corners]	(0, 1.6) rectangle (0.5, 2.1);
		\node		at	(0.25, 1.85) {\scriptsize $c$};
		\draw[<-]	(-0.5, 1.85) -- (0, 1.85);
		\draw[<-]	(0.5, 1.85) -- (1, 1.85);

		\draw[->]	(0.25, 1.1) -- (0.25, 1.6);
		\draw[->]	(0.25, 2.1) -- (0.25, 2.6);

		\draw[rounded corners]	(1.6, 1.6) rectangle (2.1, 2.1);
		\node		at	(1.85, 1.85) {\scriptsize $d$};
		\draw[<-]	(1.1, 1.85) -- (1.6, 1.85);
		\draw[<-]	(2.1, 1.85) -- (2.6, 1.85);

		\draw[<-]	(1.85, 1.1) -- (1.85, 1.6);
		\draw[<-]	(1.85, 2.1) -- (1.85, 2.6);
	
%
%

	\end{tikzpicture}
	\end{center}
	\caption{Controller four different orientations \label{fig:orientation}}
\end{figure}

%% file: evaluation.tex
\newcommand{\phiack}{\phi_{\mathsf{ack}}}
\newcommand{\phitime}{\phi_{\mathsf{time}}}

\newcommand{\NE}{\ensuremath{\mathsf{NE}}\xspace}
\newcommand{\SW}{\ensuremath{\mathsf{SW}}\xspace}

\newcommand{\SWNEx}{\ensuremath{\SW \rightarrow \NE(x)}\xspace}
\newcommand{\SWNEy}{\ensuremath{\SW \rightarrow \NE(y)}\xspace}

\newcommand{\NESWx}{\ensuremath{\NE \rightarrow \SW(x)}\xspace}
\newcommand{\NESWy}{\ensuremath{\NE \rightarrow \SW(y)}\xspace}

\section{Formal Evaluation}
\label{sec:evaluation}

This section describes the encoding of the routing protocol in the input
language of the \texttt{UPPAAL SMC} model checker and its
subsequent evaluation.  \texttt{UPPAAL SMC} is the statistical extension of
\texttt{UPPAAL}, a model checker for real-time systems represented by
networks of timed automata~\cite{UPPAAL}.  The reasons for the selection of the tool to
carry out the formal evaluation of the protocols here considered are
threefold. First, our design is associated with dense time behaviour
and requirements. Second,  \texttt{UPPAAL} implements statistical reasoning about 
properties of timed systems. Given the large state space generated by the
models, statistical model checking enables the derivation of results for
larger networks than if we had used standard model checking. Second, it
supports basic data structures expressed in the syntax of the \verb!C!
programming language, thereby allowing for concise encodings of the system's
features, e.g buffers.

\subsection{\texttt{UPPAAL SMC} Models}

The modelling here presented admits the following assumptions. First, the
network  is a $10 \times 10$ grid (as discussed in the future work section
 parameterised model checking techniques are envisaged to enable the
effective verification of larger
models~\cite{Bloem+15}). Second, in line
with the intended operation of the system, the models account only for the
routing of configuration sequences and not of arbitrary sequences of
packets.  Finally, given that nodes are identical (thus have the same
speed) and are operating very fast, we assume the presence of a global
clock and we assume that at every tick of the clock  
all nodes that may fire a transition will fire one
transition. Following the manufacturing of the first prototype chip, timing
measurements (in the form of time bounds for each operation) 
will be provided and encoded in the model in order to obtain a
more precise timing analysis.

Table~\ref{table:variants} summarises the system variants that have been
considered during the lifetime of the iteration-$0$ design process.
 The \texttt{basic} variant is as described above and assumes
a single gateway at the south-west corner of the grid. As we show below,
the \texttt{basic} system exhibits deadlocks in routing  
configuration sequences. Consequently, alternative designs had to be
evaluated so as to ``eliminate'' the deadlocks while limiting the time
requirements of the routing scheme. In particular, the \texttt{parallel}
variant assumes  that nodes are 
equipped with a different processing unit per output. More precisely,
this option is implemented in the presence of buffers within the nodes.
The buffers are used to store messages received at a node until they
are forwarded on the appropriate output, as per their destination node
and the XY algorithm. Note, however, that such sending may fail if 
the recipient node is not ready to receive (e.g. because its buffer
is already full). While in the \texttt{basic} mode the sending node will
be forced to remain idle and to retry
sending the message in the next time unit, in the \texttt{parallel}
mode, and assuming there exist further messages in its buffer, the node
will attempt to send a message on its other output channel, assuming that
such a message exists. Note that this mode was implemented in order to
explore the design possibility of implementing two independent circuits within 
a controller chip, one per output channel.

The
\texttt{acks-NE} variant includes a second gateway taking input from the
north-east corner of the network where the acknowledgements are routed as
per the XY routing algorithm (see the topology in Fig.~2). Intuitively,
this is expected to limit the congestion
emerging from routing acknowledgements from north-east coordinates to
south-west ones and data packets from south-west coordinates to
north-east ones in the \texttt{basic} variant. Note that this design
choice is also feasible given that multiple tiles, each with its own gateway,
are expected to be interconnected in the final metasurface, offering
the possibility of connecting multiple tiles to the same gateway.
 Finally the \texttt{queue-X} variant
implements a queue of size~$X$ for every node in the network.

\begin{table}
\centering
\begin{tabular}{cccc}
	\toprule
	Variant & Acknowledgements & Parallel Processing & Queue
	size\\
	\midrule
	\texttt{basic}  & \SW & no & 0 \\ 
	\texttt{parallel} & \SW & yes & 1 (to model parallelism) \\ 
	\texttt{acks-NE} & \NE & no & 0 \\
	\texttt{queue-X} & \SW & no & $X$ \\
	\bottomrule
\end{tabular}\vspace{0.2in}
\label{table:variants}
\caption{System variants}
\end{table}

All system variants are given by the parallel composition of~$100$
 timed automata modelling the nodes, and a timed automaton (automata,
respectively) representing the gateway (gateways, respectively).  The
communication between the nodes is encoded by means of four-dimensional
adjacency matrices of pairwise communication channels, where item
$[x][y][x'][y']$
denotes the communication channel
taking input from node $(x,y)$ and outputting to  node $(x',y')$. 

Fig.~\ref{fig:state_machine} depicts the timed automaton modelling the
nodes. The automaton is composed of two states (locations) and ten
transitions. Initially a node  is in  state {\it idle}. On the receipt
of a message from either input {\it in1} or {\it in2} ($\mathit{input}_1$, $\mathit{input}_2$
in Fig.~1(b)), the node goes to state
{\it Processing}. The state models the processing of the data of the packet
before the latter is routed to its destination. Whilst in this state, a
node may perform either one of the following actions: (i) if it is not the
destination node, then it can route the packet
to one of its neighbours according to the XY algorithm;
 (ii) if it is the destination node, then it will create and
route an acknowledgement to one of its neighbours towards a gateway
(either in the south-west or the north-east corner depending on the 
mode of the experiment); (iii) if it is equipped
with buffers, then it may receive a second packet which it enqueues in its buffer.  In the
figure every transition is guarded by a boolean condition determining
whether or not the transition can be fired. The condition requires from the
sender-receiver pair to respect the XY routing scheme and from the receiver
to be in a state where the packet can be queued. Further 
conditions guarding the transitions enable the synchronous evolution of the
system. Specifically a node can perform an action only when its local clock
is equal to~1; following the action, the node resets its clock; if there is
no enabled action the node simply resets its clock
whenever this equals~1. 

\input{fig-state_machine}

The timed automata modelling the gateways are responsible for generating
configuration sequences and for 
receiving the acknowledgements sent by the nodes. Following 
the topology of the network, different orderings of the packets in a
configuration sequence may induce different settings for  deadlocks and 
time requirements in routing the sequence. We therefore
consider the following configuration sequences generated by the  gateway:

\begin{enumerate}
	\item	\SWNEx. The packets are sent row by row  from
		south to north, and the packets in a row are sent from
		west to east.
	\item	\SWNEy. The packets are sent column by column
		from west to east, and the packets in a column are sent
		from south to  north.
	\item	\NESWx. The packets are sent row by row from 
		north to south, and the packets in a row are sent  from
		east to west.
	\item	\NESWy. The packets are sent 
		column by column from east to west, and the packets in a
		column are sent from north to south.
	\item	$\NE \leftrightarrow \SW$. The packets are sent alternating
		between the \SWNEx and \NESWx
		orderings at every packet sent. 
\end{enumerate}	

Indeed, as we show below, the commitment to certain orderings can enable the
implementation of simple, deadlock-free designs by building smart gateways. 

\subsection{Evaluation}

We report the experimental results obtained by checking the system variants
against specifications pertaining to deadlock-freedom and efficiency in
routing configurations sequences:
\begin{eqnarray*}
	\phiack		&\triangleq \mathsf{E}[\leq 300; 1000] \mathsf{(max:acks)}
	\\
	\phitime	&\triangleq \mathsf{E}[\leq 300; 1000] \mathsf{(max:time)} 
\end{eqnarray*}
Above, $\mathsf{acks}$ is a variable representing the number of acknowledgements that
have been received whereas $\mathsf{time}$ is a variable expressing the time taken
for all acknowledgements to be received. $\phiack$ gives the
expected maximum value of $\mathsf{acks}$ whereas $\phitime$ determines the expected
maximum value of $time$. These are calculated on the first~$300$ time units,
where empirical evaluation showed this to be an upper bound for the completion
of the protocol, and for~$1000$ traces.  During the
lifespan of the iteration-$0$ design phase, the specifications were evaluated
on progressively more complicated designs so as to derive the simplest one for
which $\phiack$ is maximised and $\phitime$ is minimised.

\input{table-results}

Table~\ref{table:exp} summarises the results obtained.
The cells with colour demonstrate the cases where
not all acknowledgements where received at the gateway,
thus the case where a deadlock is present. Note that
the times acquired in case of a deadlock 
include the deadlock traces and are thus irrelevant.
 
Evidently, the \texttt{basic} model exhibits deadlocks
under all of the configuration sequence orderings.
%
Fig.~\ref{fig:deadlock} (left) shows an \texttt{UPPAAL}-generated
simulation trace showcasing a deadlock for the \SWNEx
ordering. In the figure, node~$(0,2)$ is trying to route a data
packet to node~$(1,3)$ through node~$(0,3)$, which in turn is
trying to route an acknowledgement packet to node~$(0,1)$ through
node~$(0,3)$. Consequently node~$(0,2)$ is waiting on node~$(0,3)$ and
node~$(0,3)$ is waiting on node~$(0,2)$, thereby creating a deadlock.  

The inclusion of queue structures in the nodes may  eliminate
 deadlocks. Interestingly, to achieve this, different sizes of
queues are required for different configuration sequence orderings, ranging
from size~$1$ for the \SWNEx and $\NE \leftrightarrow \SW$
orderings, to size~$6$ for the \NESWy ordering.
Furthermore, the routing of packets under the former orderings is more
efficient. The use of parallel processing can also help to overcome
deadlocks, but only in cases \SWNEx, \NESWx,
while allowing for more efficient routing in the said cases.

The routing of the acknowledgements to a second gateway attached to
the norh-east corner of the network can
also help alleviate the deadlocks in the \NESWx, \NESWy and $\NE \leftrightarrow \SW$
orderings by, intuitively, reducing the congestion near the SW gateway.
In the other cases, adding a queue of size~$1$ is sufficient to prohibit
deadlocks from occurring. Given that the size of the queues required is
smaller than the corresponding cases with only one gateway, routing in
the presence of two gateways appears to be more efficient.

Since the gateways are cheaper than designing and implementing
queue systems and/or parallel processing capabilities, the above
experimental results suggest the design of a system with two gateways
as preferable for the purposes of the project.
Moreover, the second gateway design offers additional flexibility
and is compatible with the intended design of connecting tiles at
the gateway level to form larger metasurfaces.

A point of interest regarding the \texttt{acks-NE} design 
is the nature of the deadlock as illustrated in Table~3. 
%
Fig.~\ref{fig:deadlock} (right)
shows a part of an \texttt{UPPAAL}-generated
simulation trace that demonstrates the deadlock in a $4 \times 4$
size grid.
The problem arises when a configuration packet is routed towards
controller~$(3, 1)$, as shown with red colour. The packet necessarily
needs to be routed through controller~$(3, 2)$, which is connected to
the acknowledgement gateway. Also, in the problematic trace it happens that the configuration
packet is interleaved with acknowledgement packets, as shown with green colour,
that are routed towards controller~$(3, 2)$. The interleaving creates an input/output
dependency between controllers~$(2, 1)$, $(2, 2)$, $(3, 2)$, 
and $(3, 1)$.
Further experimentation revealed that the presence of deadlocks in the 
\texttt{acks-NE} design is due to similar cyclical dependencies among
four interconnected controllers, where acknowledgement packets and configuration
packets towards different destinations are interleaved.

Note, however, that  deadlocks are  removed when adding a queue of size~$1$.
Moreover, further experiments carried out for different grid sizes and various configuration-sequence
orderings confirmed the absence of deadlock with such a queue.
Intuitively, this can be understood as follows:
A queue allows for storing the interleaved packets to the
receivers buffer and proceed by processing the next packet that will
be sent to a different destination. In a set of nodes associated with
a circular dependency, there exists at least one node (in Fig.~\ref{fig:deadlock} (right) 
node $(3,1)$) that cannot receive input on both of its edges. Thus,
the buffer of this node will enable to break the circular dependency and
allow the flow of packets along the cycle. 
For instance, in the example of
Fig.~\ref{fig:deadlock} (right) the configuration packet from controller $(3, 2)$ to 
controller $(3, 1)$ can be stored in  the queue
of controller $(3, 1)$, thus breaking the circular dependency.

\input{fig-deadlock}

%% file: fig-state_machine.tex
\begin{figure}
	\centering
	\includegraphics[scale=0.35]{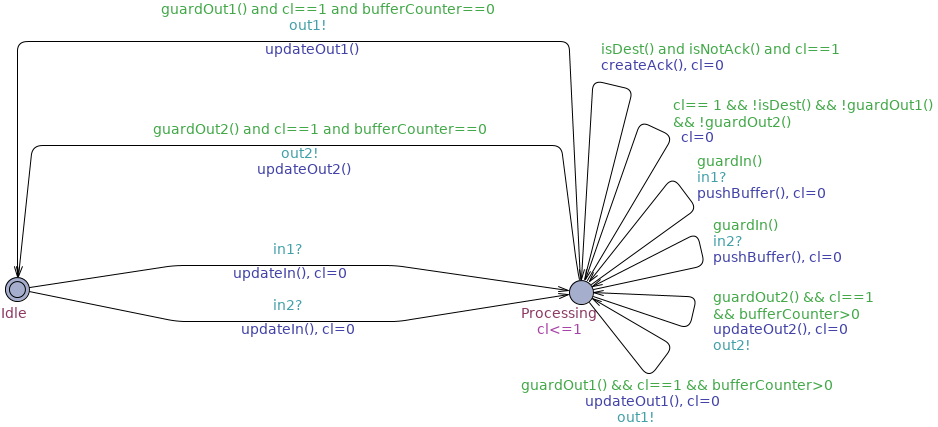}
	\caption{Timed automaton for Intra-tile Controller}

	\label{fig:state_machine}
\end{figure}

%% file: table-results.tex
\newcommand{\ackckolor}{green!25}
\newcommand{\timecolor}{red!25}

\begin{table}[t]
\centering
\begin{tabular}{|c|c|c|c|}
	\hline
	{\bf Order} & {\bf System variant} & $\boldsymbol{\phiack}$ &
	$\boldsymbol{\phitime}$ \\
	\hline
	\hline
	\multirow{4}{*}{\SWNEx}
		& \texttt{basic}
		& \cellcolor{\ackckolor} $40.85 \pm 1.23$
		& \cellcolor{\timecolor} $299.78 \pm 0.17$
	\\
		\cline{2-4}
		& \texttt{queue-1}
		& 100
		& $216.7 \pm 0.35$
	\\
		\cline{2-4}
		& \texttt{parallel}
		& 100
		& $211.46 \pm 0.34$
	\\
		\cline{2-4}
		& \texttt{acks-NE}
		& \cellcolor{\ackckolor} $99.14 \pm 0.27$
		& \cellcolor{\timecolor} $226.598 \pm 0.79$
	\\
		\cline{2-4}
		& \texttt{acks-NE-queue-1}
		& 100
		& $201.37 \pm 0.22$
	\\
	\hline
	\hline
	\multirow{4}{*}{\SWNEy} & \texttt{basic}
		& \cellcolor{\ackckolor} $2.77 \pm 0.05$
		& \cellcolor{\timecolor} 300
	\\
		\cline{2-4}
		& \texttt{queue-5}
		& 100
		& $243.03 \pm 0.43$
	\\
		\cline{2-4}
		& \texttt{parallel}
		& \cellcolor{\ackckolor} $98.74 \pm 0.68$
		& \cellcolor{\timecolor} $244.98 \pm 0.57$
	\\
		\cline{2-4}
		& \texttt{acks-NE}
		& \cellcolor{\ackckolor} $97.82 \pm 0.22$
		& \cellcolor{\timecolor} $267.38 \pm 0.24$
	\\
		\cline{2-4}
		& \texttt{acks-NE-queue-1}
		& 100
		& $213.38 \pm 0.19$  \\
	\hline
	\hline
	\multirow{4}{*}{\NESWx}
		& \texttt{basic}
		& \cellcolor{\ackckolor} $94.27 \pm 0.72$
		& \cellcolor{\timecolor} $258.49 \pm 1.43$
	\\
		\cline{2-4}
		& \texttt{queue-5}
		& 100
		& $259.54 \pm 0.34$
	\\
		\cline{2-4}
		& \texttt{parallel}
		& 100
		& $209.19 \pm 0.31$
	\\
		\cline{2-4}
		& \texttt{acks-NE}
		& 100
		& $218.58 \pm 0.12$ \\
	\hline
	\hline
	\multirow{4}{*}{\NESWy}
		& \texttt{basic}
		& \cellcolor{\ackckolor} $15.94 \pm 0.86$
		& \cellcolor{\timecolor} 300
	\\
		\cline{2-4}
		& \texttt{queue-6}
		& 100
		& $260.92 \pm 0.31$
	\\
		\cline{2-4}
		& \texttt{parallel}
		& \cellcolor{\ackckolor} $98.06 \pm 0.84$
		& \cellcolor{\timecolor} 300
	\\
		\cline{2-4}
		& \texttt{acks-NE}
		& 100
		& $219.53 \pm 0.14$
	\\
	\hline
	\hline
	\multirow{4}{*}{$\NE \leftrightarrow \SW$}
		& \texttt{basic}
		& \cellcolor{\ackckolor} $71.65 \pm 0.94$
		& \cellcolor{\timecolor} 300
	\\
		\cline{2-4}
		& \texttt{queue-1}
		& 100
		& $216.6 \pm 0.37$
	\\
		\cline{2-4}
		& \texttt{parallel}
		& \cellcolor{\ackckolor} $89.04 \pm 0.27$
		& \cellcolor{\timecolor} 300
	\\
		\cline{2-4}
		& \texttt{acks-NE}
		& 100
		& $200.36 \pm 0.31$ \\
	\hline
\end{tabular}
\vspace{0.2in}
\caption{Experimental results.}
\label{table:exp}
\end{table}

%% file: fig-deadlock.tex
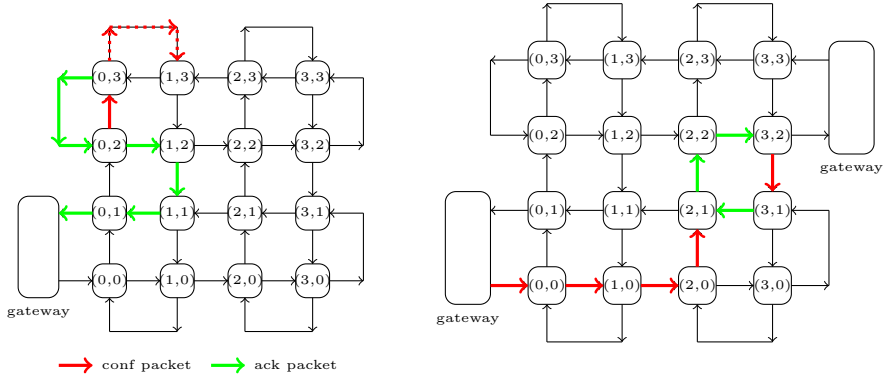
\begin{figure}
	\begin{center}
	\begin{tabular}{ccc}
		\begin{tikzpicture}[scale=0.9]
			\foreach	\x	in	{0,...,3}
				\foreach	\y	in	{0,...,3} {
					\draw[rounded corners]	(\x, \y) rectangle (\x + 0.5, \y + 0.5);
					\node	at		(\x + 0.25, \y + 0.25) {$\scriptscriptstyle (\x, \y)$};
				}

			\foreach	\x	in	{0,...,0}
				\foreach	\y	in	{0,...,1} {
					\draw[->]	(\x-0.5, 2*\y+0.25) -- (\x, 2*\y+0.25);
					\draw[<-]	(\x-0.5, 2*\y+1.25) -- (\x, 2*\y+1.25);
					\draw[->]	(2*\y+0.25, \x-0.5) -- (2*\y+0.25, \x);
					\draw[<-]	(2*\y+1.25, \x-0.5) -- (2*\y+1.25, \x);
			}

			\foreach	\x	in	{1,...,4}
				\foreach	\y	in	{0,...,1} {
					\draw[->]	(\x-0.5, 2*\y+0.25) -- (\x, 2*\y+0.25);
					\draw[<-]	(\x-0.5, 2*\y+1.25) -- (\x, 2*\y+1.25);
					\draw[->]	(2*\y+0.25, \x-0.5) -- (2*\y+0.25, \x);
					\draw[<-]	(2*\y+1.25, \x-0.5) -- (2*\y+1.25, \x);
				}

			\draw		(0.25, -0.5) -- (1.25, -0.5);
			\draw		(2.25, -0.5) -- (3.25, -0.5);

			\draw		(0.25, 4) -- (1.25, 4);
			\draw		(2.25, 4) -- (3.25, 4);

			\draw		(-0.5, 2.25) -- (-0.5, 3.25);

			\draw		(4, 0.25) -- (4, 1.25);
			\draw		(4, 2.25) -- (4, 3.25);

			\draw[rounded corners]		(-1.1, 0) rectangle (-0.5, 1.5);
			\node[below]	at		(-0.8, 0)	{\tiny gateway};

			\draw[very thick, green,<-]	(-0.5, 1.25) -- (0, 1.25);
			\draw[very thick, green,<-]	(0.5, 1.25) -- (1, 1.25);
			\draw[very thick, green,<-]	(1.25, 1.5) -- (1.25, 2);
			\draw[very thick, green,->]	(0.5, 2.25) -- (1, 2.25);

			\draw[very thick, green,->]	(-0.5, 2.25) -- (0, 2.25);
			\draw[very thick, red,->]	(0.25, 2.5) -- (0.25, 3);

			\draw[very thick, green,<-]	(-0.5, 2.25) -- (-0.5, 3.25);
			\draw[very thick, green,<-]	(-0.5, 3.25) -- (0, 3.25);

			\draw[very thick, dotted, red,->]	(0.25, 3.5) -- (0.25, 4);
			\draw[very thick, dotted, red,->]	(0.25, 4) -- (1.25, 4);
			\draw[very thick, dotted, red,<-]	(1.25, 3.5) -- (1.25, 4);

			\draw[very thick, red, ->]	(-0.5, -1) -- (0, -1);
			\node at (0, -1)[right] {\tiny conf packet};

			\draw[very thick, green, ->]	(1.75, -1) -- (2.25, -1);
			\node at (2.25, -1)[right] {\tiny ack packet};

		\end{tikzpicture}
		&\qquad&
		\begin{tikzpicture}
			\foreach	\x	in	{0,...,3}
				\foreach	\y	in	{0,...,3} {
					\draw[rounded corners]	(\x, \y) rectangle (\x + 0.5, \y + 0.5);
					\node	at		(\x + 0.25, \y + 0.25) {$\scriptscriptstyle (\x, \y)$};
				}

			\foreach	\x	in	{0,...,0}
				\foreach	\y	in	{0,...,1} {
					\draw[->]	(\x-0.5, 2*\y+0.25) -- (\x, 2*\y+0.25);
					\draw[<-]	(\x-0.5, 2*\y+1.25) -- (\x, 2*\y+1.25);
					\draw[->]	(2*\y+0.25, \x-0.5) -- (2*\y+0.25, \x);
					\draw[<-]	(2*\y+1.25, \x-0.5) -- (2*\y+1.25, \x);
			}

			\foreach	\x	in	{1,...,4}
				\foreach	\y	in	{0,...,1} {
					\draw[->]	(\x-0.5, 2*\y+0.25) -- (\x, 2*\y+0.25);
					\draw[<-]	(\x-0.5, 2*\y+1.25) -- (\x, 2*\y+1.25);
					\draw[->]	(2*\y+0.25, \x-0.5) -- (2*\y+0.25, \x);
					\draw[<-]	(2*\y+1.25, \x-0.5) -- (2*\y+1.25, \x);
				}

			\draw		(0.25, -0.5) -- (1.25, -0.5);
			\draw		(2.25, -0.5) -- (3.25, -0.5);

			\draw		(0.25, 4) -- (1.25, 4);
			\draw		(2.25, 4) -- (3.25, 4);

			\draw		(-0.5, 2.25) -- (-0.5, 3.25);

			\draw		(4, 0.25) -- (4, 1.25);

			\draw[rounded corners]		(-1.1, 0) rectangle (-0.5, 1.5);
			\node[below]	at		(-0.8, 0)	{\tiny gateway};

			\draw[rounded corners]		(4, 2) rectangle (4.6, 3.5);
			\node[below]	at		(4.3, 2)	{\tiny gateway};

			\draw[very thick, red,->]	(-0.5, 0.25) -- (0, 0.25);
			\draw[very thick, red,->]	(0.5, 0.25) -- (1, 0.25);
			\draw[very thick, red,->]	(1.5, 0.25) -- (2, 0.25);

			\draw[very thick, red,->]	(2.25, 0.5) -- (2.25, 1);
			\draw[very thick, green,->]	(2.25, 1.5) -- (2.25, 2);

			\draw[very thick, green,->]	(2.5, 2.25) -- (3, 2.25);
			\draw[very thick, red,->]	(3.25, 2) -- (3.25, 1.5);

			\draw[very thick, green,->]	(3, 1.25) -- (2.5, 1.25);

			\draw	(-0.5, -1) -- (-0.5, -1);
		\end{tikzpicture}
	\end{tabular}
	\end{center}

	\caption{
		Left: Trace showcasing a deadlock for the \texttt{basic} system under the \SWNEx configuration sequence ordering.
		Right: Trace showcasing a deadlock for the \texttt{acks-NE} system under the \SWNEx configuration sequence ordering.
		\label{fig:deadlock}
	}

\end{figure}

%% file: conclusions.tex
	\section{Conclusions and Future Work}

	The formal analysis here presented provided partial guarantees and useful insights on  
	the behaviour of the protocols and have driven their development. 
	These were used in iterations between designing the
	Hypersurface and verifying its specifications. The 
	formal evaluation was complemented  through extensive simulations via
	a simulator specifically built in the context of the project to support the
	protocol evaluation. It is worth mentioning that the  formal evaluation was
	able to pinpoint problems in instances of the model that were not discovered
	by the simulator (though they were verified by it) and, additionally, the formal
	approach had the advantage of building models and versions of the algorithm
	 much faster than  implementing them within the simulator. 
	
	However, a number of obstacles
	were encountered in the process of analysing the Hypersurface. To begin with, one of the main 
	bottlenecks was that of time. Indeed even in the context of statistical model checking,
	analysis of values required a non-negligible time: our experiments took
	up to ten minutes when run on a cluster of 12 dual-core CPUs with 24GB RAM,  and this only for $1000$ simulations (which 
	by experimentation we concluded provides an acceptable estimation of the properties
	in question). Furthermore, also relating to the state-space explosion problem
	is the fact that we have to limit our analysis for specific configuration sequences,
	though in principle it would be useful to check algorithm correctness for 
	arbitrary configuration sequences. Finally, the analysis of the results, in 
	the cases where they highlighted problems in the  execution of
	the algorithm, were difficult to interpret. Thus, in order to extract
	deadlocks in  problematic models, it was necessary to devise
	additional queries which we run by standard model checking. In this
	respect, it would be useful if the tool could be directed to store specific
	traces during the analysis. 
	
	As future work, there are various directions to explore. In the context
	of the VISORSURF project, our efforts will continue to improve the design
	of the algorithms and extend the models with more details (e.g.
	timing information). At the same time, as the analysis metrics are
	being developed, further analysis will be carried out to confirm that
	the network complies to more detailed specifications.
	
	In addition, as we have already pointed out, due to
	the state-space explosion problem our analysis is
	restricted by the size of the network and the packet configuration
	sequences.	
	To alleviate this shortcoming, sophisticated state-space
	reduction techniques need to be developed, thereby enabling the
	effective verification of the Hypersurface. In particular we will
	develop parameterised model checking techniques that enable conclusions
	to be drawn {\em irrespectively} of the size of the
	network~\cite{Bloem+15}. Specifically we believe the networks will
	admit {\em cutoffs} expressing the number of nodes that is sufficient
	to consider in order to conclude correctness for any number of nodes~\cite{KouvarosLomuscio16a,Clarke+04b}.

	Finally, the HSF design needs not only to be shown correct but also
	{\em robust} against adverse functioning conditions. Thus, we intend
	to analyse the behaviour of our design under
	various fault models and extend our routing protocols to fault-tolerant versions,
	as needed.